\documentstyle{article}\begin{document}\centerline{\bf  A two-parameter generalization of the complete elliptic integral of the second kind}\vskip .3in

\centerline{M. L. Glasser}

\centerline{Department of Physics and Department of Mathematics and Computer Science}

\centerline{Clarkson University}

\centerline{ Potsdam, NY 13699-5820 (USA)}\vskip .3in

\centerline{\bf Abstract}
\begin{quote} The double integral $$E(a,b)=\int_0^{\pi}dx\int_0^{\pi}dy\sqrt{1+a\cos\; x +b\cos\; y}\eqno(1)$$
is evaluated in terms of complete elliptic integrals. 
\end{quote}

\newpage.

V. B\^arsan[1]  has recently investigated a double integral equivalent to
$$E(a,b)=\int_0^{\pi}dx\int_0^{\pi}dy\sqrt{1+a\cos\; x+b\cos\; y}\eqno(1)$$
which, by means of an ingenious procedure,  he expressed as a derivative of a hypergeometric function of two variables. In this note we show that  (1) can be reduced in a relatively direct manner to a simple combination of complete elliptic integrals.
let us assume that $0\le a+b\le1$ and  initially that $Re\; s>0$. Then,  in the usual way one has
$$I=\int_0^{\pi}dx\int_0^{\pi}dy(1+a\cos\; x+b\cos\; y)^{-s}=$$
$$\frac{1}{\Gamma(s)}\int_0^{\infty}dst^{s-1}e^{-t}\int_0^{\pi}dx\int_0^{\pi}dy e^{-at\cos\; x-bt\cos\; y}=$$
$$\frac{\pi^2}{\Gamma(s)}\int_0^{\infty}t^{s-1}e^{-t}I_0(at)I_0(bt)dt.\eqno(2)$$
The latter is a tabulated Laplace transform[2] yielding
$$I=\pi^2F_4(s/2,(s+1)/2;1,1;u(1-v),v(1-u))\eqno(3)$$
with
$$u=\frac{1}{2}[1+a^2-b^2-\sqrt{(1+a^2-b^2)^2-4a^2}]$$
$$v=\frac{1}{2}[1-a^2+b^2-\sqrt{(1-a^2+b^2)^2-4b^2}].\eqno(4)$$
(Note that $u(1-v)=a^2$, $v(1-u)=b^2$).
Now, by analytic continuation we can take $s=-1/2$.
Next, by L. Slater's reduction formula[3], one has
$$E(a,b)=
\pi^2[\;_2F_1(-1/4,1/4;1;u)\;_2F_1(-1/4,1/4;1;v)+$$
$$\frac{1}{16}uv\;_2F_1(3/4,5/4;2;u)\;_2F_1(3/4,5/4;2;v)].\eqno(5)$$

Finally, since[4]
$$\;_2F_1(-1/4,1/4;1;z^2)=\frac{2}{\pi}\sqrt{1+z}{\bf{E}}(k)$$
$$\;_2F_1(3/4,5/4;2;z^2)=\frac{8}{\pi z^2\sqrt{1+z}}[{\bf{K}}(k)-(1+z){\bf{E}}(k)]\eqno(6)$$
$$k=\sqrt{\frac{2z}{1+z}}$$
we have the desired expression
$$\frac{1}{4}E(a,b)=2\sqrt{(1+\sqrt{u})(1+\sqrt{v})}{\bf{E}}[k(\sqrt{u})]{\bf{E}}[k(\sqrt{v})]+$$
$$\frac{{\bf{K}}[k(\sqrt{u})]{\bf{K}}[k(\sqrt{v})]}{\sqrt{(1+\sqrt{u})(1+\sqrt{v})}}-
\sqrt{\frac{1+\sqrt{u}}{1+\sqrt{v}}}{\bf{E}}[k(\sqrt{u})]{\bf{K}}[k(\sqrt{v})] -$$
$$\sqrt{\frac{1+\sqrt{v}}{1+\sqrt{u}}}{\bf{E}}[k(\sqrt{v})]{\bf{K}}[k(\sqrt{u})]\eqno(7)$$

For the case $a=b$ (3) and (7) simplify to 
$$\frac{1}{4}\int_0^{\pi}\int_0^{\pi}\sqrt{1+a(\cos\; x+\cos\; y)}dxdy=\frac{\pi^2}{4}\;_3F_2(-1/4,1/4,1/2;1,1;4a^2)=$$
$$2(1+\sqrt{u}){\bf{E}}^2(k)+(1+\sqrt{u})^{-1}{\bf{K}}^2(k)-2{\bf{E}}(k){\bf{K}}(k),\eqno(8)$$
where $u=(1-\sqrt{1-4a^2})/2$, $k=\sqrt{2\sqrt{u}/(1+\sqrt{u})}.$

\newpage
\noindent
{\bf References}\vskip .1in

\noindent
[1] V. B\^arsan, arXiv:0708.2325v1

\noindent
[2]. A.P.Prudnikov, Yu. A. Brychkov and O.I. Marichev, {\it  Integrals and Series, Vol.2.} Nauka, Moscow 1981.

\noindent
[3]  L. Slater, {\it Generalized Hypergeometric Functions}, Cambridge University Press (1966).

\noindent
[4]  A.P. prudnikov, Yu. A. Brychkov and O.I. Marichev, {\it Integrals and Series, Vol. 3.) Nauka, Moscow 1981.

\end{document}